\documentclass[a4paper, 10 pt, final]{ieeeconf}
\usepackage{amsfonts}
\usepackage{amsmath}
\usepackage{amssymb}
\usepackage{booktabs}
\usepackage{graphicx}
\usepackage{algorithm,algorithmic}
\usepackage{color}
\usepackage{cite}
\usepackage[normalem]{ulem}
\usepackage{epstopdf}

\renewcommand{\algorithmicrequire}{\textbf{Input:}}
\renewcommand{\algorithmicensure}{\textbf{Output:}}

\newtheorem{assumption}{Assumption}

\newcommand{\bY}{{\mathbb{Y}}}

\definecolor{green}{rgb}{0,0.5,0.2}

\newcommand{\bR}{{\mathbb{R}}}

\newcommand{\bU}{{\mathbb{U}}}

\IEEEoverridecommandlockouts

\title{\bf 
A Physics-Based Attack Detection Technique in Cyber-Physical Systems: A Model Predictive Control Co-Design Approach }

\author{ Mohammadreza Chamanbaz, Fabrizio Dabbene and Roland Bouffanais%
\thanks{ Mohammadreza Chamanbaz and Roland Bouffanais are with the Singapore University of Technology and Design, Singapore 487372, e-mail: ({\tt  \{chamanbaz, bouffanais\}@sutd.edu.sg})}%
\thanks{Fabrizio Dabbene is with CNR-IEIIT, Politecnico di Torino, Italy, email:({\tt fabrizio.dabbene@ieiit.cnr.it}).}
\thanks{
This work was supported in part by the National Research Foundation (NRF), Prime Minister’s Office, Singapore, under its National Cybersecurity R\&D Programme (Award No. NRF2014NCR-NCR001-040) and administered by the National Cybersecurity R\&D Directorate.
}
}

\begin{document}
\maketitle

\begin{abstract}
In this paper a novel approach to co-design controller and attack detector for nonlinear cyber-physical systems affected by false data injection (FDI) attack is proposed. We augment the model predictive controller with an additional  constraint requiring the future---in some steps ahead---trajectory of the system to remain in some time-invariant neighborhood of a properly designed reference trajectory. At any sampling time, we compare the real-time trajectory of the system with the designed reference trajectory, and construct a residual. The residual is then used in a nonparametric cumulative sum (CUSUM) anomaly detector to uncover FDI attacks on input and measurement channels. The effectiveness of the proposed approach is tested with a nonlinear model regarding level control of 
coupled tanks.


\end{abstract}

\section{Introduction}
Industrial cyber-physical systems play a crucial rule in  critical infrastructures and everyday life. Cyber-physical systems (CPSs) are constituted of physical processes (plants) communication and computation. Some examples of CPS include power grids, intelligent transportation systems, water distribution systems, aerospace systems, retail supply chain, etc. which  are all based on safety critical processes \cite{cardenas_cyber-physical_nodate}. Both cyber and physical components of a CPS are vulnerable to malicious attacks.    
Security of CPSs is of paramount importance to societies and governments. This importance has fueled a considerable research in the recent past; see survey papers   \cite{cardenas_cyber-physical_nodate,giraldo_survey_2018,ding_survey_2018,giraldo_security_2017,he_cyber-physical_2016,sridhar_cyber_2012} and references therein. In general there are two possible approaches in attack detection: i) Information Technology (IT) based methods, and ii) Physics-based (Control-based) methods. Due to the presence of physical components in CPS, the states of the system need to follow strict rules of nature, e.g. laws of physics. For instance, in a power grid, voltage of buses and current flowing through lines need to follow Kirchhoff's circuit laws. This important feature can be exploited to detect attacks in CPSs. In this paper, we focus on physics-based attack detection; hence, from now onward by attack detection, we imply physics-based attack detection.  Although attacks can be  complicated from an IT viewpoint, they tend to be rather na\"{i}ve or unsophisticated from a control prospective \cite{smith_covert_2015}. 

Most attack detection methods are observer-based techniques in which an observer is designed to estimate states of the plant. The estimated state is then compared against the actual value (measured by sensors) and a time-series residual is formed. Then, an anomaly detector is used to decide---based on the residual time series---on the presence of an adversary agent having access to the control signal and/or measurement output \cite{6859155}. The observer can be static or dynamic; for instance, the phase angle estimator in power grids is a static estimator \cite{Liu:2009:FDI:1653662.1653666}. Attack detection strategies can be broadly divided into two categories: active and passive. Passive strategies are those in which the detection mechanism does not affect the system. On the other hand, active strategies affect the control system by sending some unpredictable control commands and observing if sensors react as predicted. In \cite{7011170} a physical watermarking based detection mechanism is presented in which a random noise of known distribution is injected to the plant and a stateful anomaly detector is used to detect attacks on the system.

In most attack detectors, the control and attack detection designs  are carried out independently: the controller is designed first  and, subsequently, the detection mechanism is formulated. Contrary, in this paper we \emph{co-design} these two critical components and propose a joint control and attack detection mechanism using elements from  model predictive control (MPC). MPC is an optimization-based controller which can handle different state and control input constraints. The control signal at each sampling time is the solution of a constrained discrete-time optimal control problem \cite{mayne_model_2014,rawlings2009model,grune2017nonlinear}. For linear time-invariant systems, the resulting optimization problem is convex which can therefore be efficiently solved  using standard solvers. If the dynamical systems is nonlinear---such as the one considered in this paper, the MPC optimization problem becomes nonconvex. 

We augment the standard MPC problem with an additional constraint which restricts the future state/output trajectory to remain within some time-invariant neighborhood of a carefully designed reference trajectory. The minimization of the MPC optimization problem involves the predicted states as well as inputs over the prediction horizon. The first component of the control vector is applied to the plant and the predicted outputs are used to construct the future reference trajectory. In fact, the reference trajectory at time $k$ is the $N$th component of the predicted trajectory provided as the minimizer of the MPC problem at time $k-N$ where $N$ is the prediction horizon. The difference between actual real-time output and the reference output trajectory is stored in a residual time-series. The residual is used in  a non-parametric cumulative  sum (CUSUM) anomaly detector to decide on the presence of attack in control signal or measurement output.

\subsection{Related Literature}

While the literature on attack detection is vast (see e.g.  \cite{Bullo2013}) few works are based on a model predictive control design approach. In \cite{Barboni2018}, the authors adopt a model-based approach in order to detect cyber-attacks in a linear system equipped with a model predictive controller. The problem is formalized as a binary hypothesis test. However, the MPC structure is assumed to be given, and no co-design is considered.
A recent and interesting line of research is the one based on the design of set-theoretic receding-horizon control schemes, see for instance \cite{Lucia2016,Lucia_workshop}. In these works, a specific control architecture is designed so as to be able to detect and mitigate cyber-attacks affecting CPSs. 
In \cite{zhu_2014}, an idea based on the concept of a receding-horizon control law is presented to mitigate replay attacks. In particular, stability is proved under some  assumptions on the horizon length and the attack duration. The idea can also be used with false data injection attacks. An MPC-based attack detector using limit checking is introduced in \cite{Rosich2013}. A feasibility problem is solved online and if there does not exist a control vector being able to keep states within their safe limit during the predicted horizon, an attack is declared. We remark that all reviewed approaches are applied to linear systems, while in this paper we consider a nonlinear setup. 

\vskip 3mm

\subsection{The sequel}

The remainder of this paper is organized as follows. In Section  \ref{sec:prob formulation} we formulate the nonlinear MPC problem. The modified MPC problem along with proposed attack detection methodology are presented in Section \ref{sec:attack detection}. A numerical example of industrial cyber-physical systems is presented in Section \ref{sec:simulation} and some concluding remarks are reported in Section \ref{sec:conclusion}.

\vskip 3mm

\noindent
{\bf Notations}\\
Lowercase letters are used for vectors and uppercase ones for matrices. The symbol $X\succ0$ (resp. $X\succeq0$) is used to denote a positive (resp. positive semi-definite) matrix $X$. The set $\mathbb{N}_{>0}$ denotes the set of  positive integers. While $y_k$ denotes the measured output at time $k$, the output predicted $\ell$ steps ahead at time $k$ is denoted as $y_{\ell|k}$. We use $\mathbf{u}_{N|k}$ to denote the sequence of length $N$ of vectors $u_{0|k},\ldots,u_{N-1|k}$; the same notation is used for output vector $\mathbf{y}_{N|k}$. The Minkowski sum of $A$ and $B$ is denoted by $A\oplus B = \{a+b|a\in A ,b\in B\}$.   

\vskip 3mm

\section{Problem Formulation}\label{sec:prob formulation}

%
Consider a nonlinear cyber-physical system (CPS) whose dynamics is governed by the nonlinear discrete-time  system 
\begin{subequations}\label{eq:system no disturbance}
\begin{align}\label{eq:system no disturbance state}
	x_{k+1} =& f(x_k,(u_k+u^a_k)),\\ \label{eq:system no disturbance output}
	y_k = & x_k + y^a_k,
\end{align}
\end{subequations}
where $x_k\in\mathbb{R}^n$ is the state vector of the system at sampling time instant $k$, $u_k\in\mathbb{R}^m$ is the control signal to be applied to the system, $u^a_k\in\mathbb{R}^m$ represents an attack signal applied to the input at time $k$, 
$f:\bR^n\times\bR^m\rightarrow\bR^n$ is a nonlinear map which assigns to a state vector $x_k$ and control vector $u_k$ the successor state $x_{k+1}$, $y_k$ is the measurement signal at time $k$, and $y^a_k$ is the attack signal at the output. We assume that all states are available for feedback. The controller is spatially distributed and  the channels between controller-actuators and sensors-controller  are established by some communication network, e.g. Internet or wireless, industrial Ethernet, Fieldbus, etc. We assume that  malicious agents can gain access to these communication channels by compromising the security protocols, and as a consequence are able to inject their desired signals $u^a_k\text{ and }y^a_k$ to the system \eqref{eq:system no disturbance}. 
%
The objective is to design a model predictive controller  for the system  \eqref{eq:system no disturbance}  to be able to detect any possible attack on the control input and measurement signals. 
In the MPC framework, we usually restrict the output vector $y_k,\forall k \in \mathbb{N}_{>0} $ to live in a set $\bY$. This can be due to safety limitations; for example, the level of liquid in  a tank needs to be within its high and low limits or furnace temperature should not exceed a predefined value. Similarly, the control signal $u_k, \forall k\in \mathbb{N}_{>0}$ is also required to remain in a set $\bU$ which is to prevent any actuator saturation. 

The desired performance of the MPC controller is granted by the appropriate design of a cost function used in the optimization problem being solved at each sampling time. Assuming that $\mathbf{0}$ is the  equilibrium point of the system---in fact if $[x_*^T,u_*^T]^T\neq\mathbf{0}$ is the equilibrium point of the system \eqref{eq:system no disturbance}, we can replace $f(x_k,u_k)$ by $(f(x_k+x_*,u_k+u_*)-x_*)$ so to have $\mathbf{0}$ as the equilibrium point---the cost function $\ell(x,u):\bY\times\bU\rightarrow\bR_{>0}$ penalizes the distance of the output $y_k$ and control input $u_k$ to the equilibrium point.  A terminal cost is added to the cost function to ensure that the  MPC policy stabilizes the system. With these ingredients in mind, we use the following finite-horizon cost to be minimized at time $k$
\begin{align}\nonumber
&\ell_N(\mathbf{y}_{N+1|k},\mathbf{u}_{N|k}) = \\ \label{eq:cost function} 
&\sum_{j=0}^{N-1}\bigg(y_{j|k}^TQ y_{j|k}+u_{j|k}^TRu_{j|k}\bigg) + V_N(y_{k+N}),
\end{align}
where $Q\in\bR^{n\times n},Q\succ0,R\in\bR^{m\times m}, R\succ0$ and $V_N(y)$ is the terminal cost designed based on the Lyapunov stability theory, see \cite[Chapter 5]{grune2017nonlinear} for further details. A terminal constraint of the form $y_{N|k} \in \bY_f$ is also included in the MPC optimization problem to guarantee stability.  The terminal set $\bY_f$ is designed to grant stability to the MPC policy. With these notations, in an attack-free scenario the MPC optimization problem being solved at time $k$ is
\begin{subequations}\label{eq:MPC Original}
	\begin{align}\label{MPC Original cost}
	\min_{\mathbf{y}_{N+1|k},\mathbf{u}_{N|k}} \qquad & \ell_N(\mathbf{y}_{N+1|k},\mathbf{u}_{N|k}) \\ \nonumber
	\text{subject to:}\qquad & x_{j+1|k} = f(x_{j|k},u_{j|k}),  x_{0|k} = x_k\\ \nonumber
	& y_k = x_k \\ \nonumber
	& y_{j|k} \in \bY, j\in [1,N] \\ \nonumber
	& u_{j|k} \in \bU, j\in[0,N-1] \\  \label{eq:MPC Original Constraints}
	& y_{N|k} \in \bY_f.
	\end{align}
\end{subequations}
Denoting the minimizer of optimization problem \eqref{eq:MPC Original} with $(y^*_{1|k}, \ldots,y^*_{N|k}, u^*_{0|k},\ldots,u^*_{N-1|k})$, the  MPC control law is $u_k = u^*_{0|k}$, meaning that only the first element of the optimal control signal is applied to the system. The formulated MPC problem \eqref{eq:MPC Original} is for an attack-free scenario, i.e. $u_k^a=y_k^a=0, \forall k\in \mathbb{N}_{>0}$. In the next section, we propose a modified MPC problem along with a detection criterion which is able to detect possible attacks on  system \eqref{eq:system no disturbance}. 

\section{Attack Detection Algorithm} \label{sec:attack detection}
We first discuss the class of attacks considered in this paper, and then present the modified MPC controller and, finally the detection procedure. Here, we specifically consider the class of False Data Injection Attacks (FDI). In an  FDI attack, an attacker augments control and measurement signals with his/her desired arbitrary data by manipulating $u^a_k$ and $y^a_k$.  
Regarding the FDI attack we make the following assumption. 
\begin{assumption}\label{assumption1}
  The attacker can modify control and measurement signals by injecting its desired signals $u^a_k,y^a_k$; however, it cannot access both control and measurement channels at the same time, i.e. $\nexists k: (u^a_k\neq {0})  \& (y^a_k\neq {0})$. 
\end{assumption}
\vspace{1 ex}
Assumption \ref{assumption1} is important because if the attacker has access to the model of the system and  is able to modify both control and measurement signals, he/she can design attack signals $u^a_k$ and $y^a_k$ such that the attack remains covert, see \cite{smith_covert_2015} for further details.


\subsection{Modified MPC Algorithm}
Attack detection algorithms reported in the literature are usually independent from the controller design process. Typically, the controller is designed first, and subsequently the detection algorithm is formulated. In this paper, however, we take a novel approach by co-designing the detection algorithm and the controller. 
To this end, at each time $k$ we consider a future reference trajectory $\tilde{y}_{j|k},\,j\in[1,N]$. This constitutes a key ingredient of our approach, and is formally defined later. Next, we add an extra constraint to the MPC optimization requiring the actual output trajectory $y_{j|k},j \in [1,N]$ to remain within a specified time-invariant neighborhoods of the reference trajectory
\begin{equation}\label{eq:proximity constraints}
y_{j|k}\in\tilde{y}_{j|k}\oplus\mathcal{E}, j\in[1,N].
\end{equation}
%
Therefore, the modified MPC problem to be solved at each time $k$ reads as
\begin{subequations}\label{eq:MPC modified}
	\begin{align}\label{MPC modified cost}
	\min_{\mathbf{y}_{N+1|k},\mathbf{u}_{N|k}} \qquad & \ell_N(\mathbf{y}_{N+1|k},\mathbf{u}_{N|k}) \\ \nonumber
	\text{subject to:}\qquad & x_{j+1|k} = f(x_{j|k},u_{j|k}+a_{j+k}),  x_{0|k} = x_k\\ \nonumber
	& y_{j|k} \in \bY, j\in [1,N] \\ \nonumber
	& y_{j|k}\in\tilde{y}_{j|k}\oplus\mathcal{E}, j\in[1,N] \\ \nonumber
	& u_{j|k} \in \bU, j\in[0,N-1] \\  \label{eq:MPC modified Constraints}
	& y_{N|k} \in \bY_f.
	\end{align}
\end{subequations}

The motivation behind forcing the real-time output to remain within some time-invariant neighborhood of a reference trajectory originates from the distributed model predictive control literature, and in particular,  \cite{farina2012distributed}. In \cite{farina2012distributed} authors consider a distributed scenario where states and control inputs of each node of the network affects the neighboring nodes. Specifically, each node of the network enforces its real-time trajectory to stay within some time-invariant neighborhood of a reference trajectory. Nodes of the network receive the predicted state and control input of their neighbors and solve their local MPC problem by relying on the predicted trajectories.
%
%
Solution of the optimization problem defined in \eqref{eq:MPC modified}, i.e. $(y^*_{1|k}, \ldots,y^*_{N|k}, u^*_{0|k},\ldots,u^*_{N-1|k})$, involves predicted outputs $(y^*_{1|k}, \ldots,y^*_{N|k})$. This information can be used as the predicted reference trajectory $\tilde{y}_{j|k}$.
In fact, the reference trajectory at time $k+N,\,\tilde{y}_{k+N}$ is the $N$th component of the predicted output trajectory provided as the minimizer of the MPC problem \eqref{eq:MPC modified} at time $k,\,y^*_{k+N|k}$
\[
\tilde{y}_{k+N} = y^*_{k+N|k}.
\]
Therefore, by solving the MPC optimization \eqref{eq:MPC modified}, we gradually construct the reference trajectory.  

The set $\mathcal{E}$ defines the closeness of the actual trajectory to the future reference trajectory. Requiring the two trajectories to be very close may result in a conservative control strategy with poor performance. There is therefore  trade-off in selecting the set $\mathcal{E}$. Selecting a very small $\mathcal{E}$ compromises the performance of the MPC control strategy while very large $\mathcal{E}$ leads to poor security.  
\subsection{Anomaly Detector}
Anomaly detectors construct a series of residuals based on which they decide on the occurrence of an attack on dynamical systems. In observer-based anomaly detectors, the residual is the difference between the actual states (or outputs) and the estimated ones. In this paper, on the contrary, we use as  residual the difference between real-time output $y_k$ and the reference output trajectory $\tilde{y}_k$. Two types of detector are usually adopted in the literature: i) stateless, and ii) stateful. In a stateless test, if the residual $r_k \doteq\|y_k-\tilde{y}_k\|$ exceeds some threshold $\gamma$, an attack is declared. Contrary, in a stateful test, which is a statistical test, a new statistic $S_k$ is constructed, which keeps track of the residuals. There are a number of stateful anomaly detectors in the literature, such as simple averaging over a time window, exponential weighted moving average and, non-parametric cumulative sum known as CUSUM. In this paper, we adopt CUSUM statistic due to its popularity and  effectiveness. The CUSUM statistic $S_k$ is defined recursively as $S_0=0$ and $S_{k+1} = \max(0,S_k+r_k-\delta)$, where $\delta$ is chosen such that it prevents the CUSUM statistic to grow constantly in an attack-free scenario. An attack is declared if $S_k$ exceeds the threshold $\gamma$. Then, the CUSUM statistic is restarted, i.e. $S_{k+1}=0$. The threshold $\gamma$ is usually selected by performing extensive simulations. In fact, a small threshold may result in frequent false positives while a large value of $\gamma$ can lead to a detecting mechanism ignoring attacks. 

\begin{algorithm}[h]
\begin{algorithmic}[1]
\caption{Control and Anomaly Detection Algorithm}
\label{alg:detection}
\STATE\algorithmicrequire{ $f,N,\mathcal{E},\gamma,\delta,y_0$}
\STATE\algorithmicensure{ {\tt attack}}\\
{\bf Initialization:}\\
\STATE Construct a feasible reference trajectory $\tilde{y}_{1|0}, \ldots,\tilde{y}_{N|0}$ and set $k=0,{\tt attack}=0,S_0=0$\\
{\bf Evolution:}\\
\WHILE{${\tt attack}==0$}{
\STATE Receive $y_k$ from the plant
\STATE $S_{k+1}=\max(0,S_k+\|y_k-\tilde{y}_k\|-\delta)$
\IF{$S_{k+1}>\gamma$}{
\STATE Set ${\tt attack}=1$
\RETURN {\tt attack}
}
\ENDIF
\STATE Solve the modified MPC problem \eqref{eq:MPC modified}

\STATE Set $\tilde{y}_{k+N}=y^*_{k+N|k}$
\STATE Set $u_k = u^*_{0|k}$ and transmit $u_k$ to the plant 
}
\ENDWHILE
\end{algorithmic}
\end{algorithm}

The control and attack detection algorithm is reported in Algorithm \ref{alg:detection}.
At each sampling time, we first receive the actual measurement $y_k$ from the plant and construct the CUSUM statistic $S_{k+1}$. If $S_{k+1}$ exceeds a carefully chosen threshold $\gamma$, an attack is declared. Next, the modified MPC problem \eqref{eq:MPC modified} is solved, the reference trajectory is constructed $\tilde{y}_{k+N}=y^*_{k+N|k}$ and, the control signal $u_k=u^*_{0|k}$ is transmitted and applied to the plant.

\section{Numerical Simulations}\label{sec:simulation}
We ran extensive simulations to check the effectiveness of the presented approach. In particular, we tested the algorithm on  a nonlinear system regarding level control of two coupled tank systems.

\subsection{Level Control of Coupled Tanks}
\begin{figure}[t]
\centering
\includegraphics[width=0.7\columnwidth]{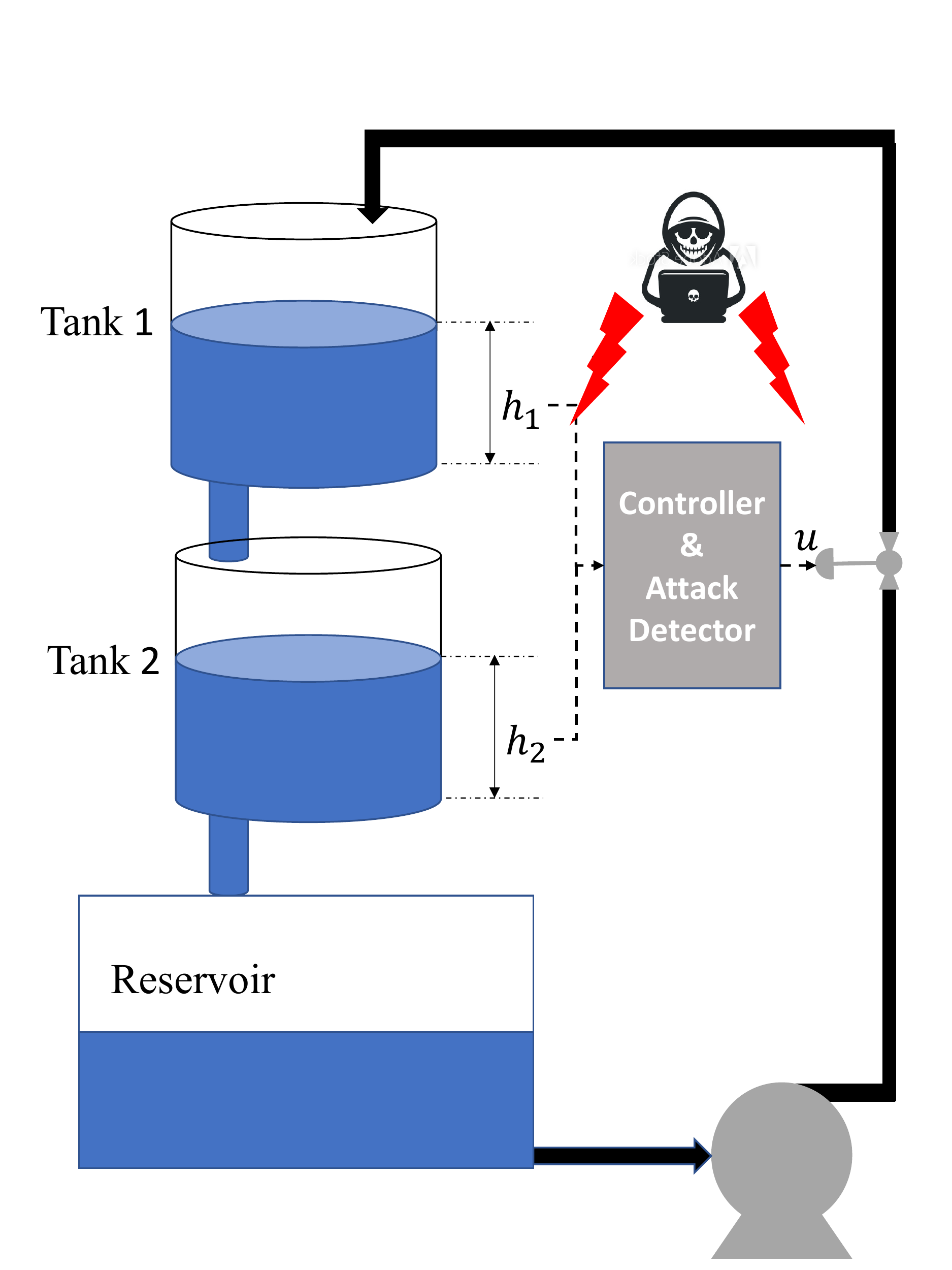}
\caption{Schematic diagram of the two coupled tanks system. The exchange of information between sensors/controller and controller/actuator are subject to false data injection attack. Levels of the two tanks are sensed and transmitted to the controller using level transmitters. The control signal is computed in the controller and then transmitted to the flow control valve to regulate the flow of liquid.}
\label{fig: Tanks}
\end{figure}

The schematic diagram of this system is shown in Fig.~\ref{fig: Tanks}. The fluid is pumped form a reservoir into the top tank. There is an opening at the bottom of the tank which allows the drainage of the fluid to the second tank. Similarly, due to the opening in the second tank, the fluid returns back into the reservoir. The goal is to control the fluid level in both tanks by manipulating the pump that connects the reservoir and Tank 1. The system dynamics can be computed using the conservation of mass and Bernoulli's equation
\begin{align}\label{eq:tanks continuous}
    \frac{dh_1}{dt} = & \frac{c_1}{\rho A_1} u - \frac{c_2}{\rho A_1}\sqrt{h_1},\\ \nonumber
    \frac{dh_2}{dt} = & \frac{c_2}{\rho A_2} \sqrt{h_1} - \frac{c_2}{\rho A_2}\sqrt{h_2},
\end{align}
where $h=[h_1,\, h_2]^T$ is the state vector containing  the level of liquid in the first and second tank respectively, $\rho$ is the density of the fluid,  $A_1, \, A_2$ are the cross-sectional area of tanks 1 and 2 respectively, $c_1,\,c_2$ are coefficients related to the tanks opening, and, $u$ is the pump rate. There are a number of ways to discretize the continuous model \eqref{eq:tanks continuous}. Here, we use the first-order forward Euler approximation to construct a discrete-time model. To this end, selecting a desired sampling time $T$, the discretized model is
\begin{align}\label{eq:tanks discrete}
    h_1(k+1) = & h_1(k)+ T\bigg( \alpha_1 u(k) - \alpha_2\sqrt{h_1(k)}\bigg),\\ \nonumber
    h_2(k+1) = & h_2(k) + T\bigg(\alpha_2 \big(\sqrt{h_1(k)}-\sqrt{h_2(k)}\big)\bigg),
\end{align}
where---assuming identical tanks---$\alpha_1 = 1.75$ and $\alpha_2 = 0.1544$.
%
%
We remark that more sophisticated discretization method such as Runge--Kutta can be used which results in a more complex discrete dynamics and consequently a more complex MPC optimization problem, see \cite[Chapter 11]{grune2017nonlinear}. The objective is to design an MPC controller to control the level of tank 2 at $0.95$. The saturation level of flow control valve is set to $1$ which means the following constraint on the control signal should hold all the time
\[
0\leq u(k) \leq 1,\quad \forall k\in \mathbb{N}_{>0}. 
\]
In order to prevent any overflow in both tanks, we require the fluid level $h(k)$ to remain below $1$. Hence, the following state constraint should be respected all the time
\[
\begin{bmatrix} 0 \\ 0 \end{bmatrix}\leq  h(k)\leq \begin{bmatrix} 1 \\ 1 \end{bmatrix}, \forall k\in\mathbb{N}_{>0}.
\]
The MPC framework formulated in Section \ref{sec:prob formulation} is for regulation problems.  To consider a tracking problem, the cost needs to be modified as 


\[
\ell_N(\mathbf{h}_{N+1|k},\mathbf{u}_{N|k}) =
\]
\[
\sum_{j=0}^{N-1}\bigg((h_{j|k}-\Bar{h}_{j|k})^TQ (h_{j|k}-\Bar{h}_{j|k})+u_{j|k}^TRu_{j|k}\bigg)
+ V_N(h_{k+N}),
\]
where $\Bar{h}_{j|k}$ is a reference  signal that output $h_{j|k}$ has to follow. 
The constraint \eqref{eq:proximity constraints} in this example is chosen to be 
\begin{equation}\label{eq:l2 norm ball}
    \|h_{j|k}-\tilde{h}_{j|k}\|_2<0.01, \quad j\in[1,N].
\end{equation}
In fact, the real-time trajectory at time $k$ is required to remain in an $\ell_2$-ball of radius $0.01$  centered at the the reference trajectory at time $k$.  Other  measures such as the infinity norm can also be used. The non-convex MPC optimization problem is formulated in YALMIP  \cite{Lofberg2004} and solved using the interior point algorithm embedded in the {\tt fmincon} solver \cite{Waltz2006}.    

\begin{figure}[t]
\centering
\includegraphics[width=0.95\columnwidth]{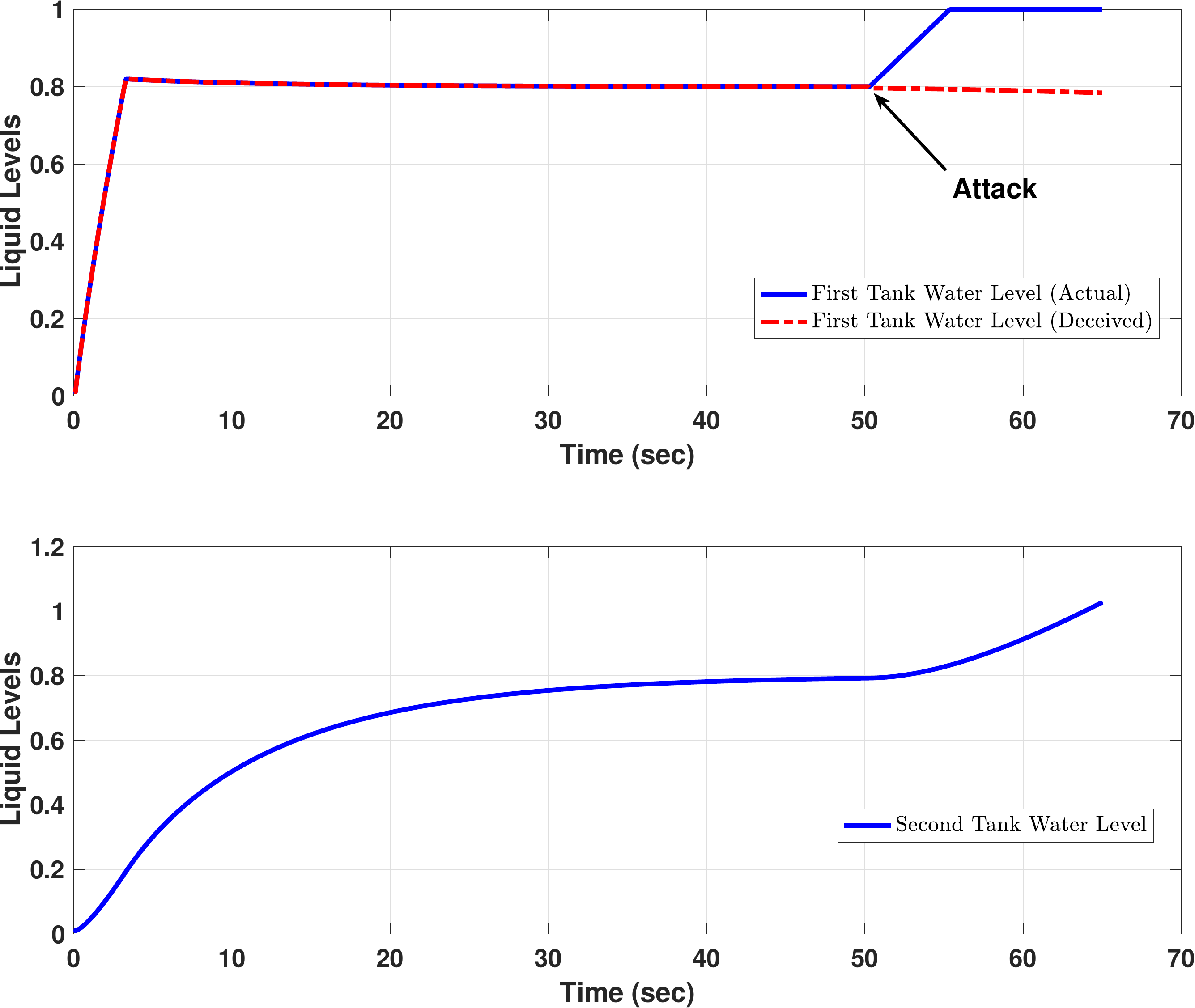}
\caption{Scenario in which  the reading form the level sensor reporting liquid level in Tank 1 is compromised starting from time $t=50$ sec. The top figure shows both the actual, as well as the deceived liquid levels in Tank 1. The bottom figure presents the liquid level at Tank 2. A simple false data injection attack causes both tanks to overflow.}
\label{fig: no detector}
\end{figure}

We first consider data injection attack on sensors. In particular, we assume that attacker has access to the level sensor installed at Tank 1. We assume that the attacker's goal is to deceive the controller to inject too much fluid in the two tanks causing them to overflow. To this end, the attacker can subtract a  positive value from the sensor reading to encourage the controller injecting more fluid into the system leading to an overflow. To show that if no detection mechanism is used, the attacker can easily lead the system to an unsafe region, we consider an FDI attack shown in Fig. \ref{fig: no detector}. Assuming that both tanks are empty at time $k=0$, i.e. $h_1(0)=h_2(0)=0$, the objective of MPC controller is to fill Tank 2 and keep its liquid level at $0.8$. We remark that at this stage the proximity constraint \eqref{eq:l2 norm ball} \emph{is not} incorporated in the MPC optimization problem. The attacker starts sending false data at time $k=500$---corresponding to $t=50$ sec---which  deceives the controller forcing it to inject more fluid into the system and eventually causes both tanks to overflow.   
This simple scenario reveals that the system without  a detection mechanism is vulnerable to FDI attacks. 

\begin{figure}[t]
\centering
\includegraphics[width=0.95\columnwidth]{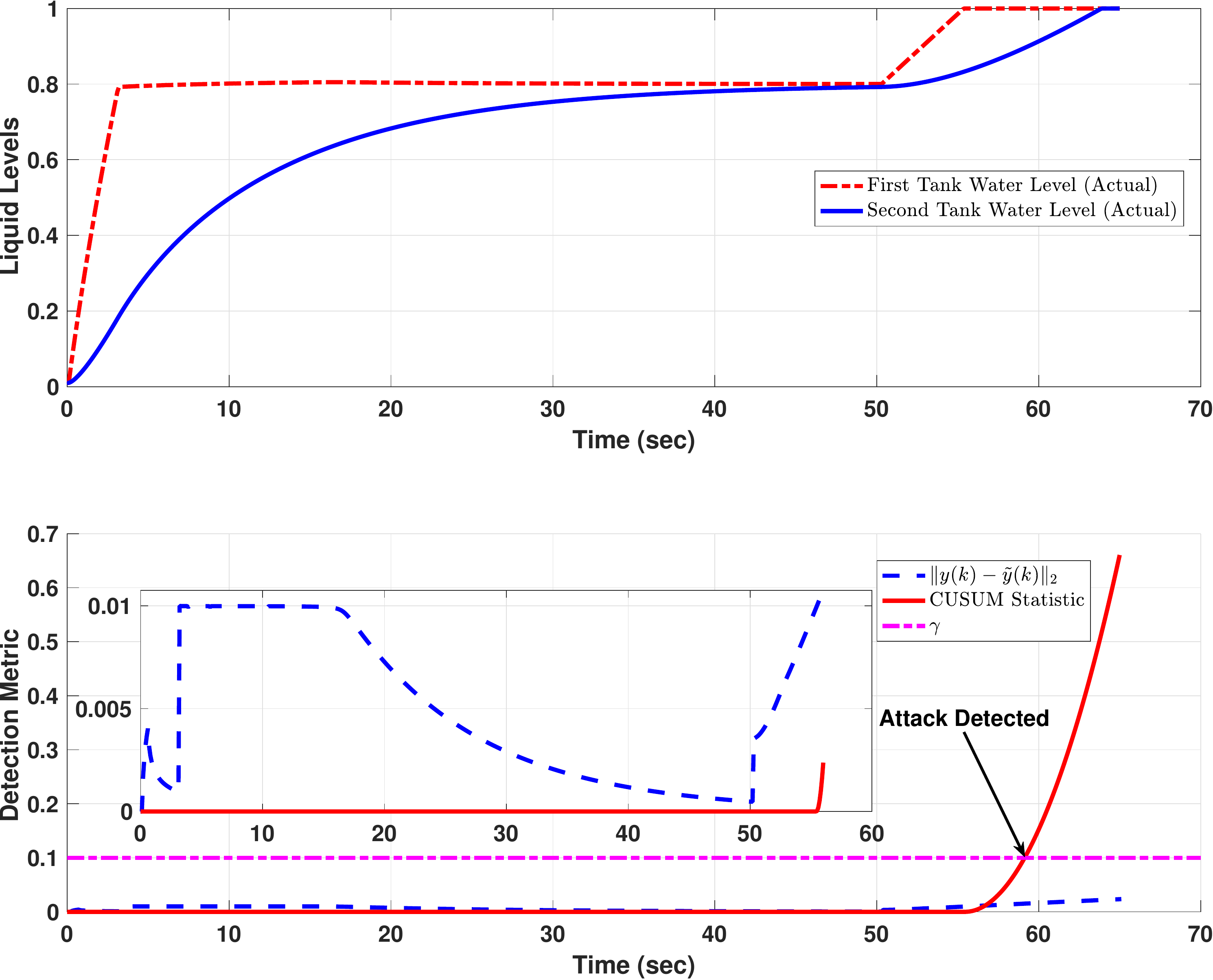}
\caption{Scenario in which  the reading from the level sensor reporting liquid level in Tank 1 is compromised starting from time $t=50$ sec. The top figure shows the actual level of both tanks. The bottom figure shows the residual $\|y(k)-\tilde{y}(k)\|_2$ as well as CUSUM statistic. Once the CUSUM statistic exceeds $0.1$, the attack is detected in the system. The inset shows a zoomed-in view of both residuals and CUSUM from $t=0$ to $t=5.6$ sec. It is clear that the residual does not exceed  $\gamma=0.01$ meaning that constraint \eqref{eq:l2 norm ball} is respected in the absence of attack.}
\label{fig: attack in y}
\end{figure}

To show the effectiveness of the proposed approach, we add the proximity constraint \eqref{eq:l2 norm ball} to the MPC optimization problem and use the anomaly detector reported in Section~\ref{sec:attack detection} to detect the attack. In particular, we use the CUSUM anomaly detector 
\[
S_0=0,\,S_{k+1} = \max(0,S_k+\|y(k)-\tilde{y}(k)\|_2-0.01),
\]
where $y_k$ is the (deceived) sensors measurement at time $k$ and $\tilde{y}_k$ is the predicted value of the levels at time $k$. Figure \ref{fig: attack in y} shows both residual $\|y(k)-\tilde{y}(k)\|_2$ as well as CUSUM statistic. 
We remark that the  same FDI attack scenario as the one reported in Fig. \ref{fig: no detector} is used to evaluate the effectiveness of the detection mechanism. The detection threshold---the parameter $\gamma$ in Algorithm \ref{alg:detection}---is selected to be $0.1$. As it can be seen from the inset in Fig. \ref{fig: attack in y}, the residual $\|y(k)-\tilde{y}(k)\|_2$ remains bellow $0.01$ meaning that the proximity constraint \eqref{eq:l2 norm ball} is respected in the absence of any attack. Starting from $t=50$ sec, the residual starts to increase resulting in an increase in the CUSUM statistic. At $t=5.92$ sec, the CUSUM statistic exceeds $0.1$ and the FDI attack is detected.

\begin{figure}[t]
\centering
\includegraphics[width=0.95\columnwidth]{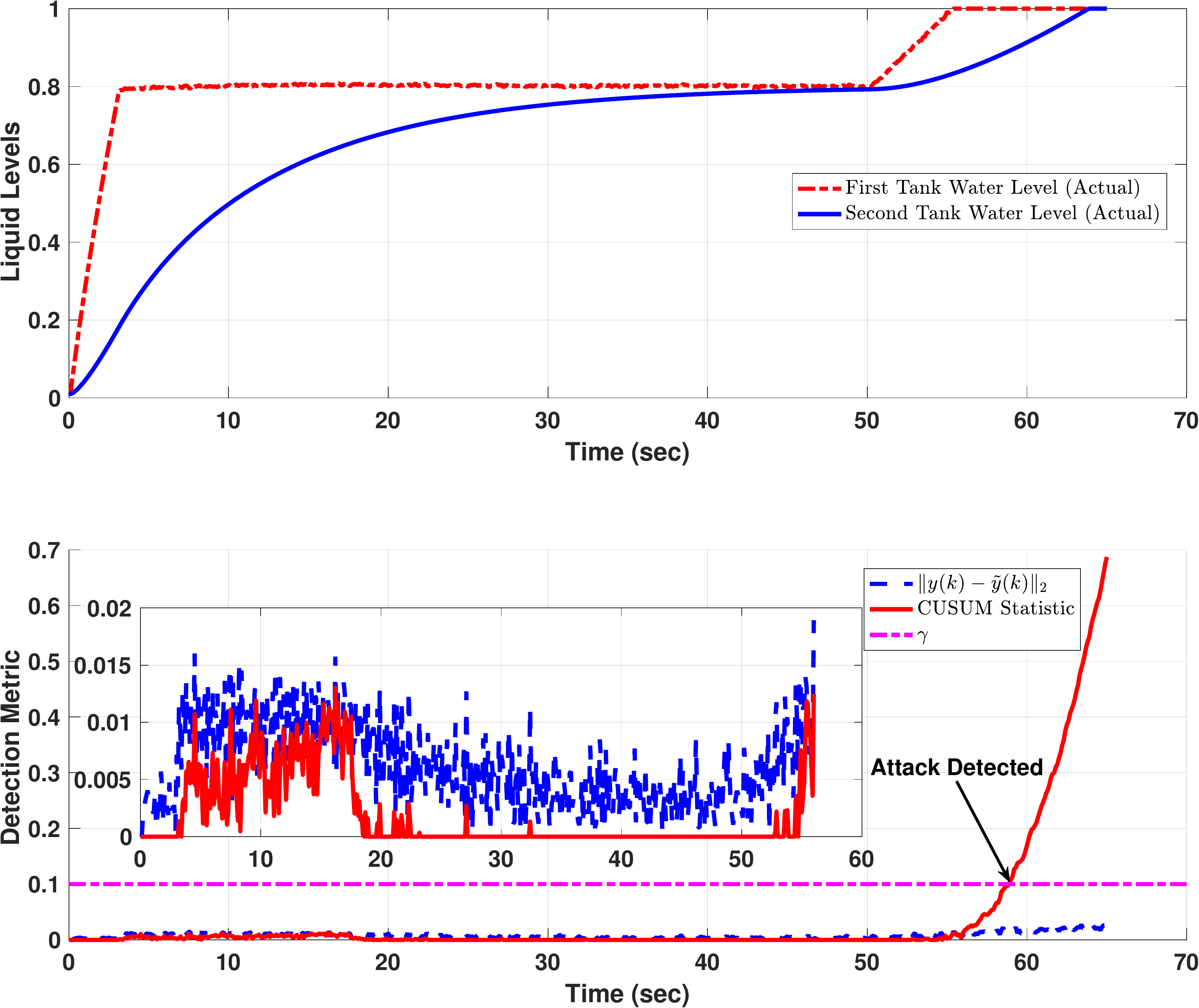}
\caption{Same scenario as the one in Fig. \ref{fig: attack in y}, but in the presence of added white Gaussian noise. The performance of the controller/detector is not affected by noise.}
\label{fig: attack in y noise}
\end{figure}

Any sensor measurement inevitably carries some noise. To see the effect of noise on the performance of the proposed detection mechanism, we add a white Gaussian noise with zero mean and $0.002$ standard deviation to the sensor measurement from both Tanks 1 and 2. The simulation reported in Fig. \ref{fig: attack in y noise} shows that the presence of noise does not have a negative effect on the performance, hence confirming the robustness to noise of the proposed approach.  

\begin{figure}
\centering
\includegraphics[width=0.95\columnwidth]{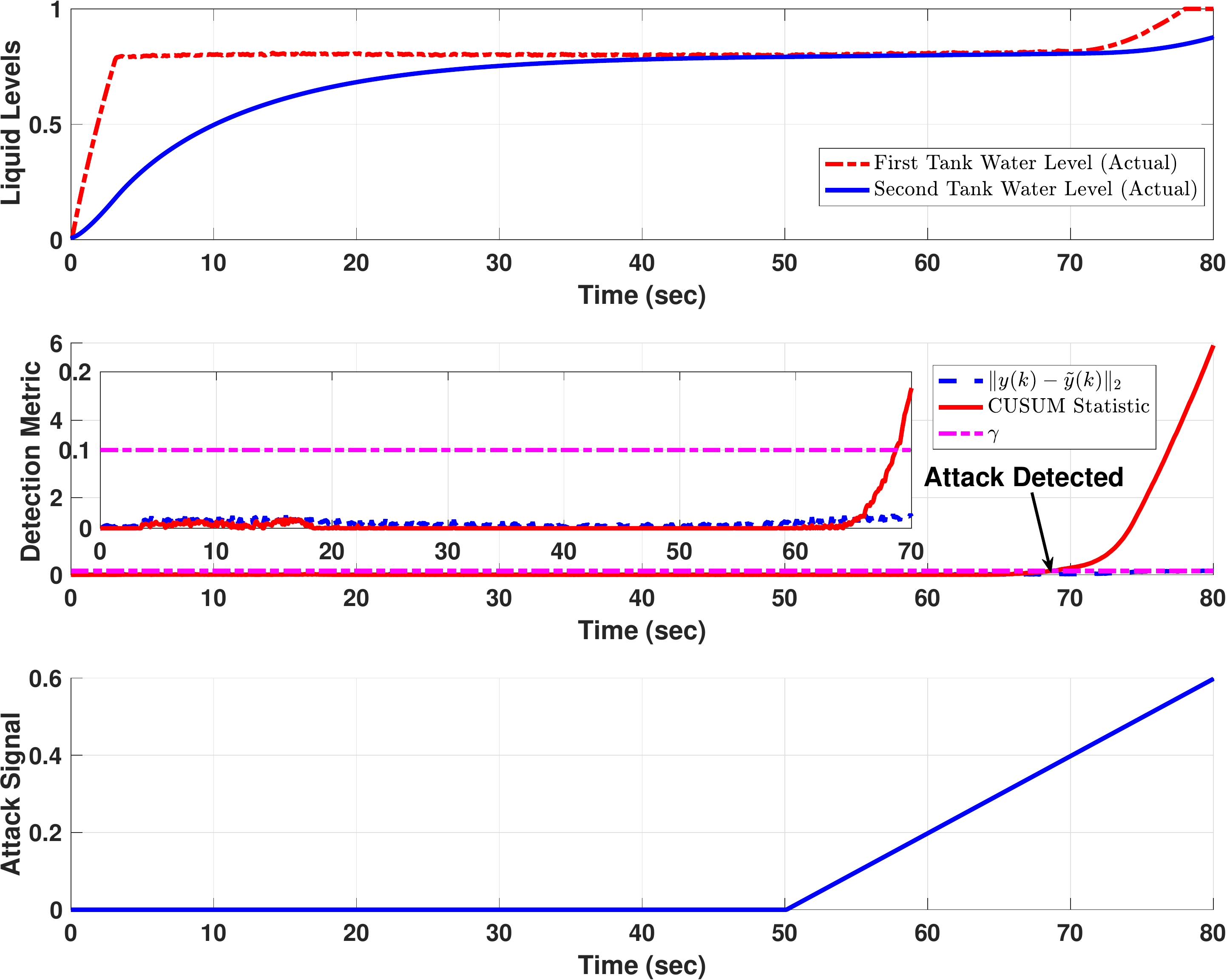}
\caption{Scenario in which the control input signal is compromised. In particular, the actual control signal $u(k)$ is replaced with $u(k)+u^a(k)$ where $u^a(k)$ is the attack signal showed in  the bottom graph. The attack signal is started at $t=50$ sec and detected at $t=6.9$ sec.}
\label{fig: attack in u}
\end{figure}

We next consider an FDI attack on the control input $u$. A simulation reporting an attack on the control signal is shown in Fig. \ref{fig: attack in u}. The attacker compromises the control signal applied to the system and augment the actual control signal $u(k)$ with an attack signal $u^a(k)$---shown in the bottom graph in Fig. \ref{fig: attack in u}. We remark that a smaller attack signal cannot bring the system to an unsafe region and hence is automatically rejected using the MPC controller.  Specifically, if the attack signal on the control input is small, the MPC controller is able to reject it by manipulating its designed control signal $u(k)$ and hence the attack signal will not have an adverse effect on the performance of the controlled system.

\section{Conclusion}\label{sec:conclusion}
An approach to co-design a controller controlling the system and an attack detector to detect false data injection attacks in control inputs and measurements is reported in this paper. 
In a model predictive controller framework, we consider a nonlinear system and require the future trajectory of the outputs to remain in some time-invariant neighborhood of a reference trajectory. Deviation of the real-time trajectory form the reference trajectory---at any point in time---is considered as a residual and used in a non-parametric cumulative sum (CUSUM) anomaly detector to detect attacks.  

Future research considers extending the proposed approach to distributed industrial control systems.

\bibliographystyle{plain}

\begin{thebibliography}{10}
	
	\bibitem{6859155}
	C.~{Bai} and V.~{Gupta}.
	\newblock On kalman filtering in the presence of a compromised sensor:
	Fundamental performance bounds.
	\newblock In {\em 2014 American Control Conference}, pages 3029--3034, 2014.
	
	\bibitem{Barboni2018}
	A.~Barboni, F.~Boem, and T.~Parisini.
	\newblock Model-based detection of cyber-attacks in networked {MPC}-based
	control systems.
	\newblock In {\em 10th IFAC Symposium on Fault Detection, Supervision and
		Safety for Technical Processes SAFEPROCESS 2018}, pages 963 -- 968, 2018.
	
	\bibitem{cardenas_cyber-physical_nodate}
	A.~Cardenas.
	\newblock Cyber-physical systems security knowledge area.
	\newblock In {\em The {Cyber} {Security} {Body} {Of} {Knowledge} (cybok)}.
	2019.
	
	\bibitem{ding_survey_2018}
	D.~Ding, Q.L. Han, Y.~Xiang, X.~Ge, and X.M. Zhang.
	\newblock A survey on security control and attack detection for industrial
	cyber-physical systems.
	\newblock {\em Neurocomputing}, 275:1674--1683, January 2018.
	
	\bibitem{farina2012distributed}
	M.~Farina and R.~Scattolini.
	\newblock Distributed predictive control: a non-cooperative algorithm with
	neighbor-to-neighbor communication for linear systems.
	\newblock {\em Automatica}, 48(6):1088--1096, 2012.
	
	\bibitem{giraldo_security_2017}
	J.~Giraldo, E.~Sarkar, A.~A. Cardenas, M.~Maniatakos, and M.~Kantarcioglu.
	\newblock Security and {Privacy} in {Cyber}-{Physical} {Systems}: {A} {Survey}
	of {Surveys}.
	\newblock {\em IEEE Design Test}, 34(4):7--17, 2017.
	
	\bibitem{giraldo_survey_2018}
	J.~Giraldo, D.~Urbina, A.~Cardenas, J.~Valente, M.~Faisal, J.~Ruths, N.O.
	Tippenhauer, H.~Sandberg, and R.~Candell.
	\newblock A {Survey} of {Physics}-{Based} {Attack} {Detection} in
	{Cyber}-{Physical} {Systems}.
	\newblock {\em ACM Comput. Surv.}, 51(4):76:1--76:36, 2018.
	
	\bibitem{grune2017nonlinear}
	Lars Gr{\"u}ne and J{\"u}rgen Pannek.
	\newblock Nonlinear model predictive control.
	\newblock In {\em Nonlinear Model Predictive Control}, pages 45--69. Springer,
	2017.
	
	\bibitem{he_cyber-physical_2016}
	H.~He and J.~Yan.
	\newblock Cyber-physical attacks and defences in the smart grid: a survey.
	\newblock {\em IET Cyber-Physical Systems: Theory \&amp; Applications},
	1(1):13--27, December 2016.
	
	\bibitem{Liu:2009:FDI:1653662.1653666}
	Y.~Liu, P.~Ning, and M.K. Reiter.
	\newblock False data injection attacks against state estimation in electric
	power grids.
	\newblock In {\em Proceedings of the 16th ACM Conference on Computer and
		Communications Security}, CCS '09, pages 21--32, New York, NY, USA, 2009.
	ACM.
	
	\bibitem{Lofberg2004}
	J.~L{\"{o}}fberg.
	\newblock Yalmip : A toolbox for modeling and optimization in matlab.
	\newblock In {\em In Proceedings of the CACSD Conference}, Taipei, Taiwan,
	2004.
	
	\bibitem{Lucia2016}
	W.~Lucia, B.~Sinopoli, and G.~Franze.
	\newblock Networked constrained cyber-physical systems subject to malicious
	attacks: a resilient set-theoretic control approach.
	\newblock In {\em arXiv preprint -- arXiv:1603.07984}, 2016.
	
	\bibitem{Lucia_workshop}
	W.~{Lucia}, B.~{Sinopoli}, and G.~{Franze}.
	\newblock A set-theoretic approach for secure and resilient control of
	cyber-physical systems subject to false data injection attacks.
	\newblock In {\em 2016 Science of Security for Cyber-Physical Systems Workshop
		(SOSCYPS)}, pages 1--5, 2016.
	
	\bibitem{mayne_model_2014}
	D.Q. Mayne.
	\newblock Model predictive control: {Recent} developments and future promise.
	\newblock {\em Automatica}, 50(12):2967--2986, December 2014.
	
	\bibitem{7011170}
	Y.~{Mo}, S.~{Weerakkody}, and B.~{Sinopoli}.
	\newblock Physical authentication of control systems: Designing watermarked
	control inputs to detect counterfeit sensor outputs.
	\newblock {\em IEEE Control Systems Magazine}, 35(1):93--109, Feb 2015.
	
	\bibitem{Bullo2013}
	F.~Pasqualetti, F.~Dorfler, and F.~Bullo.
	\newblock Attack detection and identification in cyber-physical systems.
	\newblock {\em IEEE Transaction on Automatic Control}, 58(11):2715--2729, 2013.
	
	\bibitem{rawlings2009model}
	J.B. Rawlings.
	\newblock {\em Model predictive control: Theory and design}.
	
	\bibitem{Rosich2013}
	A.~{Rosich}, H.~{Voos}, Y.~{Li}, and M.~{Darouach}.
	\newblock A model predictive approach for cyber-attack detection and mitigation
	in control systems.
	\newblock In {\em 52nd IEEE Conference on Decision and Control}, pages
	6621--6626, 2013.
	
	\bibitem{smith_covert_2015}
	R.S. Smith.
	\newblock Covert {Misappropriation} of {Networked} {Control} {Systems}:
	{Presenting} a {Feedback} {Structure}.
	\newblock {\em IEEE Control Systems}, 35(1):82--92, 2015.
	
	\bibitem{sridhar_cyber_2012}
	S.~Sridhar, A.~Hahn, and M.~Govindarasu.
	\newblock Cyber-{Physical} {System} {Security} for the {Electric} {Power}
	{Grid}.
	\newblock {\em Proceedings of the IEEE}, 100(1):210--224, 2012.
	
	\bibitem{Waltz2006}
	R.A. Waltz, J.L. Morales, J.~Nocedal, and D.~Orban.
	\newblock An interior algorithm for nonlinear optimization that combines line
	search and trust region steps.
	\newblock {\em Mathematical Programming}, 107(3):391--408, Jul 2006.
	
	\bibitem{zhu_2014}
	M.~{Zhu} and S.~{Martínez}.
	\newblock On the performance analysis of resilient networked control systems
	under replay attacks.
	\newblock {\em IEEE Transactions on Automatic Control}, 59(3):804--808, 2014.
	
\end{thebibliography}

\end{document}